\documentclass[preprint,aps,a4paper,superscriptaddress,preprintnumbers,showpacs,tightenlines]{revtex4} 
\usepackage{amsmath}
\usepackage{graphicx}
\usepackage{epsfig}

\newcommand{\mev}{\ensuremath{\mathrm{MeV}}}
\newcommand{\gev}{\ensuremath{\mathrm{GeV}}}
\newcommand{\cm}{\ensuremath{\mathrm{cm}}}
\newcommand{\ifb}{\ensuremath{\mathrm{fb}^{-1}}}
\newcommand{\stat}{\ensuremath{\mathrm{(stat.)}}}
\newcommand{\syst}{\ensuremath{\mathrm{(syst.)}}}
\newcommand{\br}{\ensuremath{\mathcal{B}}}
\newcommand{\thel}{\ensuremath{\theta_{\text{hel}}}}
\newcommand{\like}{\ensuremath{\mathcal{L}}}
\newcommand{\lrat}{\ensuremath{\mathcal{R}}}

\begin{document}

\preprint{\vbox{ \hbox{Belle Preprint 2003-24}
                 \hbox{KEK   Preprint 2003-75}
                }}

\title{Observation of the Radiative Decay $D^0 \to \phi \gamma$}

\begin{abstract}
	We report the observation of the decay
	$D^{0} \rightarrow \phi \gamma$ with a statistical significance of $5.4\sigma$
	in $78.1\,\ifb$ of data collected by the Belle experiment
	at the KEKB $e^+ e^-$ collider.
	This is the first observation of a 
 flavor-changing radiative
 decay of a charmed meson.	
	The Cabibbo- and color-suppressed decays $D^0 \to \phi \pi^0$, $\phi \eta$ are also observed for the first time.
	We measure branching fractions 
	$\br(D^{0} \rightarrow \phi \gamma) 
		= \left[ 2.60^{+0.70}_{-0.61}\,\stat\, {}^{+0.15}_{-0.17}\,\syst 
		  \right] \times 10^{-5}$, 
    	$\br(D^{0} \rightarrow \phi \pi^{0}) 
		= \left[ 8.01 \pm 0.26\,\stat \pm 0.47\,\syst
		  \right] \times 10^{-4}$,
    and 
	$\br(D^{0} \rightarrow \phi \eta) 
		= \left[ 1.48 \pm 0.47\,\stat \pm 0.09\,\syst
		  \right] \times 10^{-4}$. 
\end{abstract}


\affiliation{Budker Institute of Nuclear Physics, Novosibirsk}
\affiliation{Chiba University, Chiba}
\affiliation{University of Cincinnati, Cincinnati, Ohio 45221}
\affiliation{University of Frankfurt, Frankfurt}
\affiliation{University of Hawaii, Honolulu, Hawaii 96822}
\affiliation{High Energy Accelerator Research Organization (KEK), Tsukuba}
\affiliation{Hiroshima Institute of Technology, Hiroshima}
\affiliation{Institute of High Energy Physics, Chinese Academy of Sciences, Beijing}
\affiliation{Institute of High Energy Physics, Vienna}
\affiliation{Institute for Theoretical and Experimental Physics, Moscow}
\affiliation{J. Stefan Institute, Ljubljana}
\affiliation{Kanagawa University, Yokohama}
\affiliation{Korea University, Seoul}
\affiliation{Swiss Federal Institute of Technology of Lausanne, EPFL, Lausanne}
\affiliation{University of Ljubljana, Ljubljana}
\affiliation{University of Maribor, Maribor}
\affiliation{University of Melbourne, Victoria}
\affiliation{Nagoya University, Nagoya}
\affiliation{Nara Women's University, Nara}
\affiliation{Department of Physics, National Taiwan University, Taipei}
\affiliation{H. Niewodniczanski Institute of Nuclear Physics, Krakow}
\affiliation{Nihon Dental College, Niigata}
\affiliation{Niigata University, Niigata}
\affiliation{Osaka City University, Osaka}
\affiliation{Osaka University, Osaka}
\affiliation{Panjab University, Chandigarh}
\affiliation{Princeton University, Princeton, New Jersey 08545}
\affiliation{RIKEN BNL Research Center, Upton, New York 11973}
\affiliation{University of Science and Technology of China, Hefei}
\affiliation{Seoul National University, Seoul}
\affiliation{Sungkyunkwan University, Suwon}
\affiliation{University of Sydney, Sydney NSW}
\affiliation{Tata Institute of Fundamental Research, Bombay}
\affiliation{Toho University, Funabashi}
\affiliation{Tohoku Gakuin University, Tagajo}
\affiliation{Tohoku University, Sendai}
\affiliation{Department of Physics, University of Tokyo, Tokyo}
\affiliation{Tokyo Institute of Technology, Tokyo}
\affiliation{Tokyo Metropolitan University, Tokyo}
\affiliation{Tokyo University of Agriculture and Technology, Tokyo}
\affiliation{University of Tsukuba, Tsukuba}
\affiliation{Virginia Polytechnic Institute and State University, Blacksburg, Virginia 24061}
\affiliation{Yokkaichi University, Yokkaichi}
\affiliation{Yonsei University, Seoul}
  \author{O.~Tajima}\affiliation{Tohoku University, Sendai} 
  \author{K.~Abe}\affiliation{High Energy Accelerator Research Organization (KEK), Tsukuba} 
  \author{K.~Abe}\affiliation{Tohoku Gakuin University, Tagajo} 
  \author{H.~Aihara}\affiliation{Department of Physics, University of Tokyo, Tokyo} 
  \author{M.~Akatsu}\affiliation{Nagoya University, Nagoya} 
  \author{V.~Aulchenko}\affiliation{Budker Institute of Nuclear Physics, Novosibirsk} 
  \author{T.~Aushev}\affiliation{Institute for Theoretical and Experimental Physics, Moscow} 
  \author{A.~M.~Bakich}\affiliation{University of Sydney, Sydney NSW} 
  \author{A.~Bay}\affiliation{Swiss Federal Institute of Technology of Lausanne, EPFL, Lausanne}
  \author{I.~Bizjak}\affiliation{J. Stefan Institute, Ljubljana} 
% ...
\collaboration{The Belle Collaboration}


\maketitle

\begin{figure}[t]
  \begin{center}
    \includegraphics[scale=0.33]{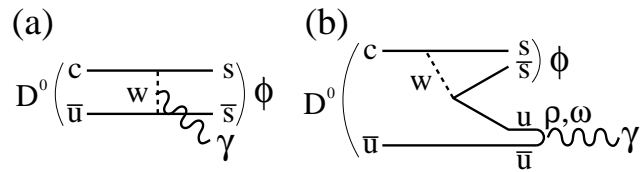}
  \end{center}
  \caption{Decay diagrams for the decay $D^0 \to \phi \gamma$:
    (a) short-distance, (b) long-distance process.
	}
  \label{diagram_eps}
\end{figure}

Flavor-changing radiative 
decays of the charmed meson system,
$D \rightarrow V \gamma$  where $V$ is a vector meson, 
have not previously been observed.
In the Standard Model, the short-distance contribution to these decays
is negligible
(the branching fraction from this contribution is predicted to be less than $10^{-8}$),
and the long-distance contribution due to a vector meson coupling to a photon
is expected to be dominant~\cite{bajc_fajfer_1995_bib, pakvasa_1995_bib}.
Examples of short- and long-distance processes for the decay 
$D^{0} \rightarrow \phi \gamma$ are shown in Figure \ref{diagram_eps}.
The branching fraction of this mode is expected to lie in the range
$(0.04 - 3.4) \times 10^{-5}$~\cite{pakvasa_1995_bib, fajfer_1997_bib};
the current 90\% confidence level (C.L.) upper limit from CLEO is
$\br(D^{0} \rightarrow \phi \gamma) < 1.9 \times 10^{-4}$~\cite{cleo_rad_decay_bib}. 
In the $B$-meson system, where the short-distance contribution to 
$B \rightarrow X_{d} \gamma$ ($X_d$ is $\rho, \omega$) decay 
can be used to measure the CKM matrix element
$V_{td}$, theoretical estimates of the long-distance contribution 
are also uncertain \cite{BtoPhoGam_bib}.
Measurement of radiative $D$ meson decays therefore provides an
important constraint on the 
interpretation of $B \rightarrow X_{d} \gamma$ results.

In this paper, we present the first observation of the
$D^{0} \rightarrow \phi \gamma$ decay.
The analysis is based on $78.1\,\ifb$ of data collected at 
the $\Upsilon(4S)$ resonance by the Belle detector~\cite{belle_detector}
at the KEKB $e^+e^-$ collider~\cite{kekb}. 
KEKB is a pair of electron storage rings with asymmetric energies,
3.5 GeV for $e^+$ and 8 GeV for $e^-$, and a single interaction point.
The Belle detector is a large-solid-angle general purpose spectrometer
that consists of a three-layer silicon vertex detector (SVD), 
a 50-layer central drift chamber (CDC), 
an array of aerogel threshold \v{C}erenkov counters (ACC),
a barrel-like arrangement of time-of-flight scintillation counters (TOF),
and an electromagnetic calorimeter comprised of CsI(Tl) crystals (ECL),
located inside a superconducting solenoid coil that provides a 1.5 T
magnetic field.
An iron flux-return located outside of the coil is instrumented to
detect $K^{0}_{L}$ mesons and to identify muons (KLM).
 
According to a Monte Carlo (MC) study, the most important backgrounds
to $D^0 \to \phi \gamma$ are the Cabibbo- and color-suppressed decays
$D^{0} \rightarrow \phi \pi^{0}$ and $\phi \eta$, which have not 
previously been observed.
We therefore conduct a search for these decay
modes as well.
To reduce the combinatorial background, 
$D^0$ candidates are combined with $\pi^+$ to form $D^{\ast+}$
candidates.
We calculate the difference in invariant mass,
$\Delta M = M_{D^{0}_{cand}\pi^+} - M_{D^{0}_{cand}}$, and
require $143.4 ~ \mev/c^2 < \Delta M < 147.4 ~ \mev/c^2$.

Charged particle tracks are reconstructed in the SVD and CDC and
required to be consistent with originating from the interaction region;
the nearest approach of the trajectory to the collision point,
which is determined run-by-run,
is required to pass
$|dr| < 0.5~\cm$ and $|dz| < 1.5~\cm$, 
where $dz$ is taken in the direction of the positron beam 
and $dr$ is in the plane perpendicular to it.
Particle identification (ID) likelihoods for the pion ($\like_{\pi}$) and kaon ($\like_{K}$) hypotheses
are determined from the ACC response, specific ionization ($dE/dx$)
measurement in the CDC, and the time-of-flight measurement for each track.
To identify kaons (pions), we apply a mode-dependent requirement on the likelihood ratio
$\lrat \equiv \like_K/(\like_K+\like_\pi)$:
a criterion of $\lrat > 0.6$ ($\lrat < 0.1$) yields an efficiency of
86\% (79\%) for kaons (pions).
The rate at which pions (kaons) are misidentified as kaons (pions) under
these criteria is 8\% (5\%).
We select $\phi$ candidates from $K^{+}K^{-}$ combinations in the invariant mass
range $1.01 ~ \gev/c^2 < M_{K^{+}K^{-}} < 1.03 ~ \gev/c^2$,
and use combinations from the mass sidebands
($\phi_{sb}$),
$0.99 ~ \gev/c^2 < M_{K^{+}K^{-}} < 1.00~ \gev/c^2$ 
and $1.04 ~ \gev/c^2 < M_{K^{+}K^{-}} < 1.05 ~ \gev/c^2$,
to estimate the background under the $\phi$ peak.

Neutral $\pi$ mesons ($\pi^{0}$) are formed by pairing photons, each with
energy $E_{\gamma} > 50\,\mev$, 
and requiring a pair invariant mass within $\pm 16\,\mev/c^2$ ($\sim 3\sigma$) of the
nominal $\pi^0$ mass; the photon momenta are then recalculated with a $\pi^0$ 
mass constraint \cite{mass_constraint_fit_bib}.
For $\eta \to \gamma\gamma$ and $D^0 \to \phi \gamma$ reconstruction, 
photon candidates passing $E_{\gamma} > 50\,\mev$ are selected 
if no pairing with any other photon in the event ($E_{\gamma} >
20\,\mev$) yields an invariant mass that is consistent with the 
$\pi^0$ mass (the $\pi^0$ veto).
Otherwise, $\eta$ meson candidates are formed from pairs of selected
photons and
required to have an invariant mass within $\pm 40\,\mev/c^2$ 
($\sim 4\sigma$) of the $\eta$ mass. 
The photon momenta are then recalculated with an $\eta$ mass
constrained fit \cite{mass_constraint_fit_bib}.

We make the following requirements on laboratory momentum or energy: $P_{\pi^0} > 750\,\mev/c$ for $D^{0} \rightarrow \phi \pi^{0}$, $P_{\eta} > 500\,\mev/c$ for $D^{0} \rightarrow \phi \eta$,
and $E_{\gamma} > 450\,\mev$ for $D^{0} \rightarrow \phi \gamma$.
The $D^{*}$ momentum in the $e^+e^-$ center-of-mass is required to satisfy
$P^{*}_{D^{*}} > 2.9\,\gev/c$ for all modes;
this criterion is optimised for the selection of $D^{0} \rightarrow \phi \gamma$ and is above the kinematic limit for $D^{*}$'s produced in $B$ decays at the $\Upsilon$(4S) resonance.
These criteria are determined using MC such that they do not introduce
bias for the signal in data.

\begin{figure}[b]
  \begin{center}
    \includegraphics[scale=0.4]{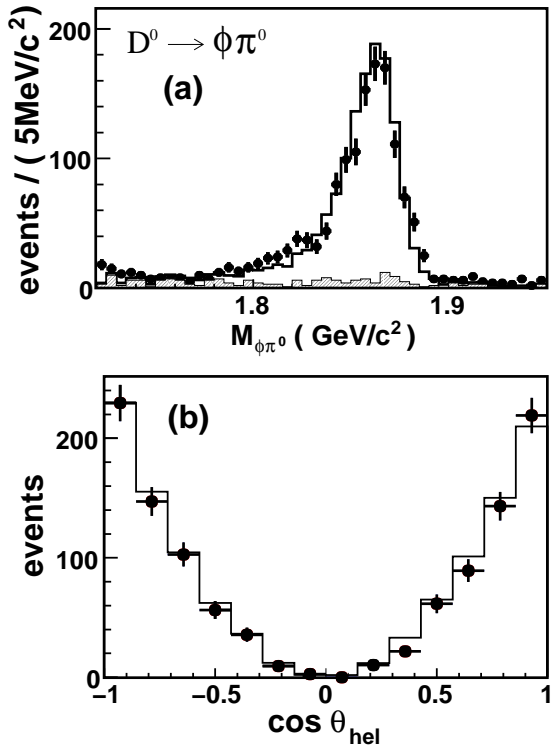}
  \end{center}
  \caption{
 	$D^0 \to \phi \pi^{0}$: 
	(a) invariant mass distribution
		for data (points with errors),
        combinations of $\phi$-sideband and $\pi^0$
        (shaded histogram), 
        and sideband + MC signal shape, 
        scaled to the result of the fit described in the text
        (solid histogram);
 	(b) fitted $D$ yield in bins of $\cos \thel$ (points with errors) 
		(\thel\ is the $\phi$ helicity angle)
		and the MC prediction (histogram).
	}
  \label{phipi0-mass_eps}
\end{figure}

The reconstructed $M_{\phi\pi^0}$ distribution shows a clear
enhancement at the $D^{0}$ mass
(Figure \ref{phipi0-mass_eps}(a)).
The distribution from non-resonant $D^{0} \rightarrow K^+K^- \pi^{0}$
decay also peaks 
at the $D^0$ mass.
Its contribution is evaluated using $\phi$-sideband $D^0$ candidates,
constructed from $\phi_{sb}\pi^0$ combinations
and found to comprise $3.1\pm 0.9$\% of the net $D^0$ yield.
The signal yield is extracted by a $\chi^2$ fit to the sideband-subtracted
$M_{\phi\pi^0}$ distribution,
assuming an exponential shape for the background and a signal shape
developed by the Crystal Ball experiment~\cite{xbal}. 
We obtain $1254 \pm 39$ events ($\chi^2/ndf=25.6/38$).
To measure the helicity state, we form the helicity angle
$\theta_{hel}$, the angle between the $K^+$ and $D^0$ 3-momenta in
the rest frame of the $\phi$ meson.
Due to the conservation of angular momentum, 
the distribution in $\cos\theta_{hel}$ is expected to be
proportional to $\cos^2\theta_{hel}$ 
for $D^{0}\rightarrow \phi \pi^{0}$ and $\phi \eta$ decays 
but proportional to $\sin^2\theta_{hel}$ 
($ = 1-\cos^2\theta_{hel}$ )
for $D^{0} \rightarrow \phi \gamma$.
The distribution for $\phi \pi^{0}$ candidates agrees well with
the MC expectation (Figure \ref{phipi0-mass_eps}(b)).
When a $\phi \pi^0$ or $\phi \eta$ decay is reconstructed as a
$\phi \gamma$ candidate by missing one photon, the distribution
of $\cos\theta_{hel}$ is still close to $\cos^2\theta_{hel}$.
The background to $D^{0} \rightarrow \phi \gamma$ from 
$\phi \pi^{0} / \phi \eta$ 
decay is thus strongly suppressed by a requirement on $\theta_{hel}$.

\begin{figure}[t]
  \begin{center}
    \includegraphics[scale=0.38]{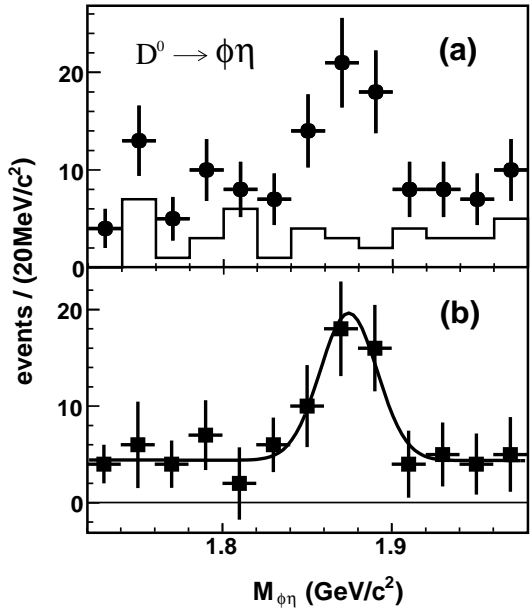}
  \end{center}
  \caption{
 	$D^0 \to \phi \eta$: 
	(a) invariant mass distribution
		for data (points with errors) and 
        combinations of $\phi$-sideband and $\eta$
        (histogram);
	(b) background-subtracted distribution,
		with a fit function described in the text (solid curve).
 	}
 \label{phieta-mass_eps}
\end{figure}
	
The $M_{\phi\eta}$ and 
$\phi$-sideband 
($M_{\phi_{sb}\eta}$)
distributions are shown in Figure \ref{phieta-mass_eps}(a).
We extract the signal yield
from the sideband-subtracted $M_{\phi\eta}$ distribution shown in Figure
\ref{phieta-mass_eps}(b). 
The $\chi^2$ fit yields $31.1 \pm 9.8$ signal events, where Gaussian and first-order polynomial shapes are assumed for
the signal and background, respectively.
The significance of the signal, taken to be $\sqrt{-2\ln\left( L_{0}/L_{best}\right) }$ where $L_{best}$ and $L_{0}$ are the maximum likelihood values with the signal floated and fixed to zero, respectively,  is $4.4\sigma$.

For the radiative mode, we require 
$| \cos\theta_{hel} | <0.4$.
A clear peak is observed at the $D^0$ mass in the $M_{\phi\gamma}$ invariant mass distribution  (Figure \ref{phigam-mass_eps}(a)).
The signal yield is extracted using the binned maximum likelihood method.
The shapes of signal and background from 
$D^{0} \rightarrow \phi \pi^{0}$, 
$\phi \eta$ and $D^{+} \rightarrow \phi \pi^{+} \pi^{0}$ (feed-down)
are obtained by MC simulation. 
The reliability of the simulation is checked by studying $K_{S}^0\gamma$
combinations with an invariant mass near the $D^0$ mass:
the process $D^0 \rightarrow K_{S}^0\gamma$ is forbidden,
so any peaking in the signal region must be
due to $D^0 \rightarrow K_{S}^0\pi^0$ and
$K_{S}^0\eta$ events,
similar to the feed-down background in this analysis.
We observe no peaking.
We find similar distributions in data and MC.
The feed-down rates are normalized to the observed rates for the source
modes, and the systematic uncertainty is due to the uncertainty on the
source rates.
The shape of the combinatorial background is estimated
by fitting a first order polynominal to 
$\phi$-sideband 
($\phi_{sb}\gamma$)
candidates.
The normalization of this contribution is allowed to float in the fit. 
The extracted yield is $27.6^{+7.4}_{-6.5}\,\stat\, {}^{+0.5}_{-1.0}\,\syst$ events. 
The significance of the signal is $5.4\sigma$.

\begin{figure}[b]
  \begin{center}
    \includegraphics[scale=0.4]{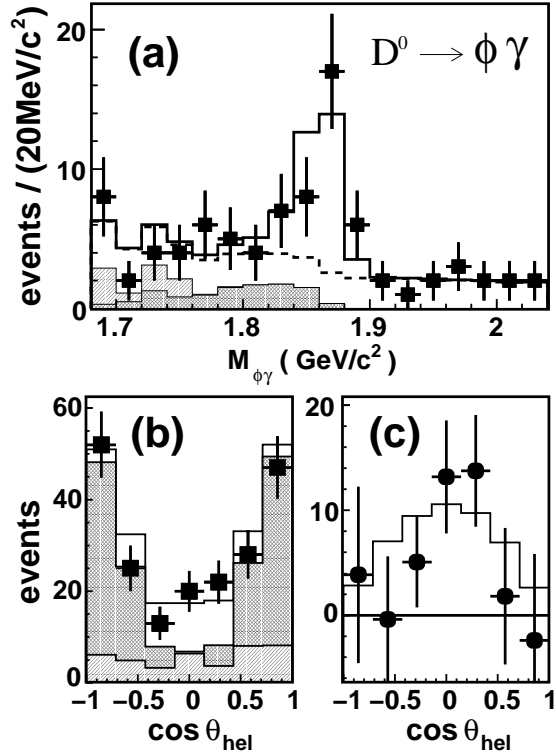}
  \end{center}
  \caption{
	$D^0 \to \phi \gamma$: 
	(a) invariant mass distribution
		for data (points with errors),
		the fit described in the text (solid histogram),
		the background component of the fit (dashed),
        $\phi\pi^{0}$ background (dark shading), 
        and the sum of $\phi\pi^{0}$, $\phi\eta$, 
        and $D^{+} \rightarrow \phi\pi^{+}\pi^{0}$ backgrounds (light);
 	(b) $\cos \thel$ distribution in the signal region,
		with the MC predictions: total (solid), total background (dark),
		and non-$\phi \pi^0$ background (light);
	(c) background-subtracted $\cos \thel$ distribution
		and the MC prediction (histogram).
  	}
  \label{phigam-mass_eps}
\end{figure}

With the signal region defined as $[1.78-1.92]\,\gev/c^2$ and the requirement on $\cos\theta_{hel}$ released, the distribution in 
$\cos\theta_{hel}$ shows an excess of signal over background near $\cos\theta_{hel} \sim 0$ (Figure \ref{phigam-mass_eps}(b)). 
It is clear that, without the requirement on helicity angle, the feed-down from $D^{0} \rightarrow \phi \pi^{0}$ and $\phi \eta$ is
large. 
Figure~\ref{phigam-mass_eps}(c) shows
the helicity angle distribution after background subtraction.
It is consistent with the $\sin^2\theta_{hel}$ distribution expected for 
the $D^{0} \rightarrow \phi \gamma$ signal.

  \begin{table}[b]
   \begin{center}
    \caption{Efficiencies for reconstructed modes [\%]}
    \label{eff-table}
    \begin{tabular}{lcccc}
     \hline
     \hline 
     Mode ($D^{0} \rightarrow$)
          & $K^{+} K^{-}$ 
          & $\phi \pi^{0}$ 
          & $\phi \eta$ 
          & $\phi \gamma$ 
          \\
     \hline
     Reconstruction
          & $10.4$
          & $6.48$
          & $2.22$
          & $4.32$
          \\
     
          &$\pm 0.1$
          &$\pm 0.04$
          &$ \pm 0.05$
          &$\pm 0.04$
\\
     $\br(\phi \rightarrow K^{+}K^{-})$
          & ---
          & 49.20
          & 49.20
          & 49.20
          \\
     $\br(\pi^{0} / \eta \rightarrow 2\gamma)$
          & ---
          & 98.80
          & 39.43
          & ---
          \\
     Efficiency correction
          & ---
          & 97.90
          & 96.20
          & 98.50
          \\
     \hspace{8pt} for $\pi^{0}$ / $\eta$ / $\gamma$ & & & & \\
     \hline
     Total
          & 10.40
          & 3.08
          & 0.413
          & 2.09
          \\
     \hline
     \hline
    \end{tabular}
    \caption{Estimated fractional systematic errors [\%]}
    \label{syst-table}
    \begin{tabular}{lccc}
     \hline
     \hline 

          & $D^{0} \rightarrow \phi \pi^{0}$ 
          & $D^{0} \rightarrow \phi \eta$ 
          & $D^{0} \rightarrow \phi \gamma$ 
          \\
     \hline
     Tracking etc.
          & 0.59
          & 0.59
          & 0.59
          \\
     Particle ID
          & 2.74
          & 2.74
          & 2.74
          \\
     $\Delta M$
          & 0.98
          & 0.98
          & 0.98
          \\
     Mass of $\phi$
          & 0.94
          & 0.94
          & 0.94
          \\
     Efficiency correction
          & 2.75
          & 1.62
          & 1.94
          \\
     \hspace{10pt} for $\pi^{0} / \eta / \gamma$ & & & \\
     Fitting \& BG
          & 1.67
          & 2.01
          & +2.46/-3.99
          \\
     $\br(D^{0} \rightarrow K^{+} K^{-})$
          & 3.40
          & 3.40
          & 3.40
          \\
     $\br(\phi \rightarrow K^{+} K^{-})$
          & 1.42
          & 1.42
          & 1.42
          \\
     $\br(\pi^{0}/\eta \rightarrow 2\gamma)$
          & 0.03
          & 0.66
          & ---
          \\
     MC statistics
          & 0.99
          & 2.76
          & 1.11
          \\
     \hline
     Total
          & 5.88
          & 6.16
          & +5.86/-6.65
          \\
     \hline
     \hline
    \end{tabular}
   \end{center}
  \end{table}
  \begin{table}
   \begin{center}
    \caption{Measured branching ratios \br$_f$/\br$_{K^+K^-}$ 
             for $D^0$ decay modes ($f$)
             and branching fractions 
             \br$_f$ = (\br$_f$/\br$_{K^+K^-}$)$\times$\br$_{K^+K^-}$.
             The first error is statistical and the second is systematic.}
    \label{result-table}
    \begin{tabular}{lll}
     \hline
     \hline 
     Mode
          & \multicolumn{1}{c}{\br$_f$/\br$_{K^+K^-}$}
          & \multicolumn{1}{c}{\br$_f$ $(\times 10^{-4})$}
          \\
     \hline 
     ~$\phi \gamma$
          & $
            \left[ 6.31\,^{+1.70}_{-1.48}\,{}^{+0.30}_{-0.36}\right]\times10^{-3}
            $
          & $
	    \left[ 2.60\,^{+0.70}_{-0.61}\,{}^{+0.15}_{-0.17}\right]\times10^{-1}
	    $
          \\
     ~$\phi \pi^0$
          & $
            \left[ 1.94 \pm 0.06 \pm 0.09 \right] \times 10^{-1}
            $
          & $
	    8.01 \pm 0.26 \pm 0.47
	    $
          \\
     ~$\phi \eta$
          & $
            \left[ 3.59 \pm 1.14 \pm 0.18 \right] \times 10^{-2}
            $
          & $
	    1.48 \pm 0.47 \pm 0.09
	    $
          \\
     \hline
     \hline
    \end{tabular}
   \end{center}
  \end{table}

To minimize systematic uncertainties,
we measure the branching fraction as a ratio to $D^{0} \rightarrow K^{+}
K^{-}$, where we have a signal of $21787\pm 226$ events, and derive the
branching fraction using the world average 
${\cal B}(D^{0} \rightarrow K^{+} K^{-}) = (4.12\pm0.14)\times 10^{-3}$
\cite{PDG_bib}.
Many of the systematic errors associated with tracking and particle ID are at least partially canceled in the ratio.

The reconstruction efficiencies are estimated via MC simulation. 
Some differences in efficiency between data and MC have been noted, in particular for photon and $\pi^0$ reconstruction.
For $\pi^0$, this is studied using the double ratio of two modes of the $\eta$,\[
   \frac{\varepsilon_{data}(2\pi^0)}{\varepsilon_{MC}(2\pi^0)}
   = 
   \frac{N_{data}(\eta \rightarrow 3 \pi^0)
   /N_{MC}(\eta \rightarrow 3\pi^0)}
   {N_{data}(\eta \rightarrow \gamma \gamma)
   /N_{MC}(\eta \rightarrow \gamma \gamma)},
\]
  where
   $
   R_{\varepsilon}(\pi^0 \rightarrow \gamma\gamma) 
   |_{\eta \rightarrow 3 \pi^0}
   = 
   R_{\varepsilon}(\eta \rightarrow \gamma\gamma)
   $ $
   (R_{\varepsilon} \equiv \varepsilon_{data}/\varepsilon_{MC})
   $
is assumed. 
The efficiency correction factors estimated from this study are found to
be $(97.9 \pm 2.7)\%$ for $\pi^0$ ($E_{\gamma} > 50\,\mev$ and
$P_{\pi^0} > 750\,\mev/c$), $(96.2 \pm 1.6)\%$ for $\eta$ ($E_{\gamma}
> 50\,\mev$ and $P_{\eta} > 500\,\mev/c$) and $(98.5 \pm 1.9)\%$ for 
a signal $\gamma$  with $E_{\gamma} > 450\,\mev$
(assuming $\varepsilon_{\pi^0} = \varepsilon_{\gamma} \times \varepsilon_{\gamma}$). The overall detection efficiencies are summarized in Table \ref{eff-table}.
The $\pi^0$ veto results in a low reconstruction efficiency 
for the $\phi \eta$ mode.
The branching fractions for the observed $D^0$ decays and their ratios to the
reference mode are summarized in the Table \ref{result-table}.

  The systematic uncertainties are summarized in Table \ref{syst-table}.
The errors derived from the fitting process are estimated by varying the fit 
range, signal shape, and bin width in the fits; errors due to the backgrounds
are estimated by varying background contributions within their	uncertainties.
The uncertainty in the acceptance due to the requirement on $M_{K^+K^-}$
for $\phi$ candidates is estimated by observing changes in the ratio 
$R_{D^0 \rightarrow \phi \pi^0} \equiv \left[ N_{data}/N_{MC} \right]_{D^0 \rightarrow \phi \pi^0}$ while shifting the signal region by $\pm 1\,\mev/c^2$. 
Other systematic errors are estimated by observing changes to the double ratio, $R_{D^0 \rightarrow \phi \pi^0} / R_{D^0 \rightarrow K^+K^-}$. 
The error due to the uncertainty in the tracking efficiency is estimated by
changing the maximum impact parameter criteria
from 0.5 cm to 0.3 cm for $|dr|$ 
and from 1.5 cm to 1.0 cm for $|dz|$.
Similarly, the errors due to
particle ID and the 
$D^{*}-D^{0}$ 
mass difference are estimated
by loosening the likelihood ratio cut to $\lrat > 0.5$ ($\lrat < 0.5$)
for kaon (pion) selection and by shifting the $\Delta M$ window by 
1 MeV/$c^2$ on each side.
Due to the difference in 
the $K^\pm$ momentum distribution between 
the $D^0\rightarrow\phi\pi^0$ and $D^0\rightarrow K^+K^-$ modes, the
uncertainty on the particle ID efficiency does not exactly cancel in the
ratio. 
The uncertainties in the branching fractions of submodes are taken from
the current world averages \cite{PDG_bib}.

To summarize, we have observed for the first time a radiative decay 
of the $D$ meson, in the mode $D^{0} \rightarrow \phi \gamma$.
  We also observe two other rare decays, 
  $D^{0} \rightarrow \phi \pi^0$ and $\phi \eta$, 
  which are Cabibbo-suppressed and color-suppressed modes 
  and constitute backgrounds for the radiative mode.
  The radiative $D \rightarrow V \gamma$ decays
  are expected to be dominated by long-distance contributions; 
however, the theoretical uncertainty on the rate is very large.
The observed rate of $D^{0} \rightarrow \phi \gamma$  constitutes
  evidence for significant long-distance contributions, 
and 
it provides an anchor for the further development of non-perturbative QCD.

We wish to thank the KEKB accelerator group for the excellent
operation of the KEKB accelerator.
We acknowledge support from the Ministry of Education,
Culture, Sports, Science, and Technology of Japan
and the Japan Society for the Promotion of Science;
the Australian Research Council
and the Australian Department of Education, Science and Training;
the National Science Foundation of China under contract No.~10175071;
the Department of Science and Technology of India;
the BK21 program of the Ministry of Education of Korea
and the CHEP SRC program of the Korea Science and Engineering Foundation;
the Polish State Committee for Scientific Research
under contract No.~2P03B 01324;
the Ministry of Science and Technology of the Russian Federation;
the Ministry of Education, Science and Sport of the Republic of Slovenia;
the National Science Council and the Ministry of Education of Taiwan;
and the U.S.\ Department of Energy.


\begin{thebibliography}{99}
\bibitem{mass_constraint_fit_bib}
The momenta of the photons are shifted while the candidate $\pi^0$ or $\eta$
mass is constrained to the nominal value, and the $\chi^2$ of the shifts
is minimized.
\end{thebibliography}
\end{document}